\documentclass{elsart}

\usepackage{amsmath,amssymb,graphicx,bm}

\newcommand{\be}{\begin{equation}}
\newcommand{\ee}{\end{equation}}
\newcommand{\bea}{\begin{eqnarray}}
\newcommand{\eea}{\end{eqnarray}}
\newcommand{\lm}{\Lambda}
\newcommand{\vlowk}{V_{{\rm low}\,k}}

\newcommand{\tr}{{\rm Tr}}

\newcommand{\fmi}{\, \text{fm}^{-1}}
\newcommand{\fmiq}{\, \text{fm}^{-3}}
\newcommand{\mev}{\, \text{MeV}}
\newcommand{\gevi}{\, \text{GeV}^{-1}}

\newcommand{\ohfnn}{\Omega_{1,{\rm NN}}}
\newcommand{\ohfnnn}{\Omega_{1,{\rm 3N}}}
\newcommand{\oa}{\Omega_{2,{\rm a}}}
\newcommand{\on}{\Omega_{2,{\rm n}}}

\newcommand{\tl}{\widetilde{l}}
\newcommand{\tlp}{\widetilde{l'}}
\newcommand{\tj}{\widetilde{J}}

\begin{document}

\begin{frontmatter}

\title{Neutron matter at finite temperature}

\author{L. Tolos}$^1$,
\ead{tolos@fias.uni-frankfurt.de}
\author{B. Friman}$^2$ and
\ead{b.friman@gsi.de}
\author{A. Schwenk}$^3$
\ead{schwenk@triumf.ca}
\address{$^1$FIAS, J.W.~Goethe Universit\"at, Ruth-Moufang-Str.~1, \\
D-60438 Frankfurt am Main, Germany \\
$^2$GSI, Planckstr.~1, D-64291 Darmstadt, Germany \\
$^3$TRIUMF, 4004 Wesbrook Mall, Vancouver, BC, Canada, V6T 2A3}

\begin{abstract}
We calculate the neutron matter equation of state at finite temperature
based on low-momentum two- and three-nucleon interactions. The free energy 
is obtained from a loop expansion around the Hartree-Fock energy, including
contributions from normal and anomalous diagrams. We focus on densities
below saturation density with temperatures $T \leqslant 10 \mev$ and
compare our results to the model-independent virial equation of state
and to variational calculations. Good agreement with the virial equation
of state is found at low density. We provide simple estimates for the
theoretical error, important for extrapolations to astrophysical conditions.
\end{abstract}

\end{frontmatter}

\section{Introduction}

The nuclear equation of state plays a central role in astrophysics,
for problems
ranging from the structure of neutron stars~\cite{LP}, neutron star
mergers~\cite{Oechslin} to core-collapse supernovae~\cite{Mezza,Janka}.
Astrophysical applications probe 
the equation of state at the extremes of isospin 
and temperature: The mass of a neutron star depends mainly on the 
equation of state of neutron matter up to densities 
$\rho \sim 4 \rho_0$~\cite{KB}, 
where $\rho_0 = 0.16 \fmiq$ is the saturation density of symmetric nuclear 
matter, while supernova explosions are most sensitive
to the properties of nucleonic matter at subnuclear densities and
MeV temperatures~\cite{Mezza}. For many regimes of interest, the equation of
state has to be extrapolated
from the conditions reached with existing and upcoming 
experimental facilities. Therefore, reliable theoretical input is needed.
In this paper, we present a study of neutron matter at 
finite temperature, as part of a program to improve the nuclear 
equation of state input for astrophysics. 

Conventional 
nucleon-nucleon (NN) interactions are nonperturbative as a result
of several sources. First, there is a strong short-range repulsion,
which leads to bound states of the ``flipped'' potential $\lambda 
V_{\rm NN}$ for small, negative $\lambda$. This implies
that $\lambda=1$ is far outside the radius of convergence.
Consequently, at least the summation of particle-particle ladder 
diagrams is required~\cite{BETHE71}. 
Second, the tensor force, which is singular at short distances, requires 
iteration in the triplet channels~\cite{tensor,Fleming}. Finally, 
there are physical bound and nearly-bound states in the S-waves, which render 
the perturbative Born series divergent. Recently, it was shown 
that the first two sources of nonperturbative behavior depend on the
choice of NN interaction, and can be removed by evolving nuclear 
forces to low-momentum interactions 
$\vlowk$~\cite{VlowkReport,VlowkPLB,VlowkRG,Vlowksmooth} with 
cutoffs around $2 \fmi$~\cite{Vlowknucmatt,VlowkWeinberg}.
An important additional advantage is that 
the corresponding leading-order three-nucleon (3N)
interactions from chiral effective field theory (EFT) become
perturbative in light nuclei for cutoffs 
$\lm \lesssim 2 \fmi$~\cite{Vlowk3N}.

At sufficient density ($\rho \gtrsim 0.01 \rho_0$~\cite{FNN} in nuclear
matter), Pauli blocking eliminates the shallow 
bound states, and thus the particle-particle channel becomes
perturbative~\cite{Vlowknucmatt}. Consequently, the Hartree-Fock
(HF) approximation
is a good starting point for low-momentum NN and 3N interactions,
and perturbation theory (in the sense of a loop expansion) around
the HF energy becomes tractable. The perturbative
character is due to a combination of Pauli blocking and an appreciable 
effective range (see also Ref.~\cite{dEFT}). The 3N interaction 
is essential for nuclear matter saturation~\cite{Vlowknucmatt}, while the
contributions to the potential energy remain compatible with 
EFT power-counting estimates. Furthermore,
the equation of state becomes significantly
less cutoff dependent with the inclusion of the dominant second-order 
contributions. In this paper, we extend the investigation of 
Ref.~\cite{Vlowknucmatt} to neutron matter at subsaturation
densities, $\rho < \rho_0$, and generalize the perturbative approach 
to finite temperature.

Based on the work of Kohn, Luttinger and Ward~\cite{KL,LW}, at 
finite temperature the loop expansion around the HF free energy 
can be realized by the perturbative expansion of the free
energy. In this paper, we include the first-order NN and 3N 
contributions, as well as anomalous and normal second-order 
diagrams with NN interactions. We defer
3N contributions beyond the HF level and
higher-order corrections to future work. The pressure, entropy and
energy are calculated using standard thermodynamic 
relations. Since low-momentum
interactions are energy independent, the Matsubara
sums can be carried out analytically.

Low-momentum interactions $\vlowk$ and the corresponding
3N forces are defined by sharp or smooth regulators 
with a variable momentum cutoff $\lm$. 
Varying the cutoff is a powerful tool to estimate the
theoretical errors due to neglected higher-order many-body
interactions and to assess the completeness of the 
calculations. We use the cutoff dependence to provide simple
error estimates, and find that the cutoff dependence is
reduced significantly, when second-order contributions are
included. The possibility of estimating theoretical errors
is an important step towards reliable extrapolations to 
astrophysical conditions. Finally, we compare our results to the
virial equation of state~\cite{vEOSnuc,vEOSneut} and to
variational calculations~\cite{FP}.
The low-density behavior is in good agreement with the 
virial equation of state. Our results for the 
energy per particle (see Fig.~\ref{energy})
highlight the importance of a correct 
finite-temperature treatment of second and higher-order 
correlations.

This paper is organized as follows. In Sect.~\ref{formalism},
we discuss the perturbative expansion at finite temperature
and give the expressions for the evaluated diagrams. 
Our results for the free energy, pressure, entropy
and energy are presented in Sect.~\ref{results}. We conclude 
and give an outlook in Sect.~\ref{conclusions}.

\section{Loop expansion at finite temperature}
\label{formalism}

\begin{figure}
\begin{center}
\includegraphics[scale=0.625,clip=]{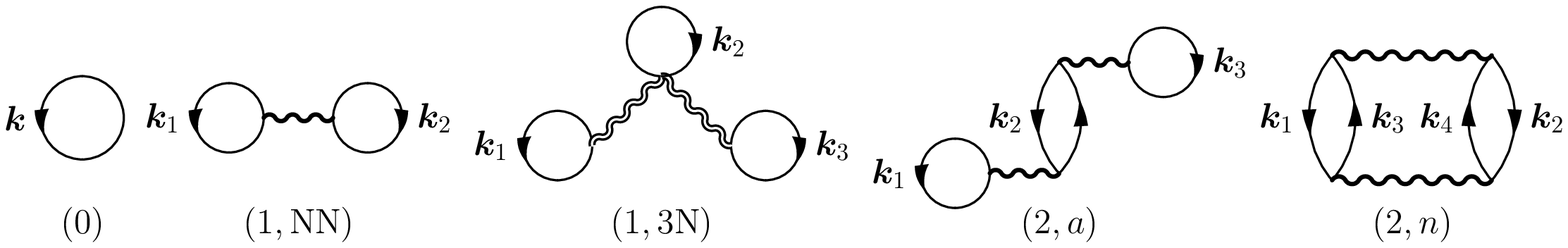}
\end{center}
\caption{Non-interacting ($0$); first-order NN ($1,{\rm NN}$) and
3N ($1,{\rm 3N}$); second-order anomalous ($2,{\rm a}$) and normal
($2,{\rm n}$) contributions to the grand-canonical potential 
$\Omega(\mu,T,V)$. The wiggly and double-wiggly lines denote
antisymmetrized low-momentum NN and 3N interactions, respectively.}
\label{diags}
\end{figure}

We consider the perturbative expansion of the grand-canonical
potential,
\be
\Omega(\mu,T,V) = - \beta \, \ln {\mathcal Z}(\mu,T,V) \,,
\ee
where ${\mathcal Z}(\mu,T,V)$ denotes the partition function
of the interacting Fermi system, $\mu$ is the chemical 
potential, $\beta=1/T$ the inverse temperature and $V$ the 
volume. We include the first-order NN and 3N contributions, 
$\ohfnn$ and $\ohfnnn$, as well as the second-order anomalous and 
normal contributions with NN interactions, $\oa$ and $\on$. 
The grand-canonical potential is then given by
\begin{align}
\Omega &= \Omega_0 + \Omega_1 + \Omega_2 + \ldots \nonumber \\[2mm]
&= \Omega_0 + ( \ohfnn + \ohfnnn ) + ( \oa + \on ) + \ldots \,,
\end{align}
where terms of the same order are enclosed in brackets, and 
$\Omega_0$ is the grand-canonical potential of the non-interacting 
system,
\be
\frac{\Omega_0}{V} = - 2 \, T
\int\frac{d{\bf k}}{(2\pi)^3} \:
\ln \bigl( 1 + e^{-\beta(\epsilon_k-\mu)} \bigr) 
= -2 \int\frac{d{\bf k}}{(2\pi)^3} \: 
\frac{k^2}{3 m} \: n_k \,.
\ee
Here, $\epsilon_k=k^2/(2m)$ is the free single-particle energy,
with $m$ the nucleon mass, and $n_k = 1/[e^{\beta(\epsilon_k
-\mu)} + 1]$ is the Fermi-Dirac distribution function. The different
contributions are depicted diagrammatically in Fig.~\ref{diags}.

The loop expansion around the HF energy is realized
by the perturbative expansion of the free energy $F(N,T,V)$, which is
obtained by a Legendre transformation of the grand-canonical potential 
with respect to the chemical potential,
\be
F(N,T,V) = \Omega(\mu,T,V) +\mu \, N \,.
\ee
The mean particle number $N \equiv \langle N \rangle$ is given by
\be
N(\mu,T,V) = - \frac{\partial \Omega}{\partial \mu} \biggr|_{T,V}
= - \frac{\partial \Omega_0}{\partial \mu} \biggr|_{T,V} 
- \frac{\partial \Omega_1}{\partial \mu} \biggr|_{T,V} 
- \frac{\partial \Omega_2}{\partial \mu} \biggr|_{T,V} - \ldots \,.
\label{eq:N}
\ee
In order to invert Eq.~(\ref{eq:N}) for the chemical potential
$\mu(N,T,V)$, we follow the treatment of Kohn and Luttinger~\cite{KL}
and expand $\mu$ to the same order
\be
\mu = \mu_0 + \mu_1 + \mu_2 + \ldots \,,
\label{eq:mu}
\ee
where the particle number is counted as order zero. The lowest
order term $\mu_0$ 
is the chemical potential of a non-interacting system
with the same density $\rho = N/V$ as the interacting system.

Expanding each term on the right-hand side of Eq.~(\ref{eq:N}) 
around $\mu=\mu_0$ and solving for
the chemical potential order by order leads to
\be
N = - \frac{\partial \Omega_0}{\partial \mu} \biggr|_{\mu_0}
\quad \text{and} \quad
\mu_1 = - \frac{\partial \Omega_1/\partial \mu}{
\partial^2 \Omega_0/\partial \mu^2} \biggr|_{\mu_0} \,,
\label{eq:mu1}
\ee
and correspondingly for the free energy
\begin{align}
F &= \Omega_0(\mu_0) 
+ (\mu_1+\mu_2) \, \frac{\partial \Omega_0}{\partial \mu} \biggr|_{\mu_0} 
+ \frac{1}{2} \, \mu_1^2 \, \frac{\partial^2 \Omega_0}{
\partial \mu^2} \biggr|_{\mu_0} 
+ \Omega_1(\mu_0) + \mu_1 \, \frac{\partial \Omega_1}{\partial \mu} 
\biggr|_{\mu_0} + \Omega_2(\mu_0) \nonumber \\[2mm]
&+ \mu_0 N + (\mu_1+\mu_2) N + \ldots \,.
\end{align}
Using Eq.~(\ref{eq:mu1}), we find
\be
F(N) = F_0(N) + \Omega_1(\mu_0) + \Omega_2(\mu_0)
- \frac{1}{2} \, \frac{(\partial \Omega_1/\partial \mu)^2}{
\partial^2 \Omega_0/\partial \mu^2} \biggr|_{\mu_0} \,,
\ee
where $F_0(N) = \Omega_0(\mu_0) + \mu_0 N$ is the free energy of 
the non-interacting system.
Since we neglect the contribution of 3N interactions in second-order
diagrams, we consistently keep only the first-order NN part $\ohfnn$
in the term $(\partial \Omega_1/\partial \mu)^2 |_{\mu_0}$.
Consequently, we have
\be
F(N) = F_0(N) + \Omega_1(\mu_0) + \on(\mu_0) +
\left[ \oa(\mu_0) - \frac{1}{2} \, \frac{(\partial \ohfnn/\partial 
\mu)^2}{\partial^2 \Omega_0/\partial \mu^2} \biggr|_{\mu_0}
\right] \,.
\label{eq:F}
\ee
According to
the Kohn-Luttinger-Ward theorem~\cite{KL,LW}, the
term in the square brackets vanishes at $T = 0$ for isotropic,
normal Fermi systems, since the anomalous diagram cancels against 
the contribution induced in $\ohfnn$ by the 
shift of the chemical potential. Thus, the
above expansion ensures that the $T \to 0$ limit is
correctly reproduced.

In the $T \to 0$ limit $\mu_0 = \epsilon_{\rm F}
= k_{\rm F}^2/(2m)$, where $k_{\rm F}$ is the Fermi momentum, and the 
ground state energy of the interacting system is given by $F 
\to E = E_0 + \Omega_1(\epsilon_{\rm F}) + \on(\epsilon_{\rm F})
+ \ldots$. For a momentum-independent
contact interaction, the square bracket in 
Eq.~(\ref{eq:F}) vanishes at
all temperatures, since in this case the HF self-energy is
momentum independent. Consequently, the thermodynamic potential derived
from the free energy Eq.~(\ref{eq:F}) corresponds exactly to the loop 
expansion around the
HF energy. For finite-range interactions, the HF self-energy is
momentum dependent, and the cancellation is exact only in the
zero-temperature limit. At finite temperature, the momentum
dependence of the HF self-energy is therefore treated perturbatively.

The pressure, entropy and energy follow from the
free energy using standard thermodynamic relations. The entropy per
particle $S/N$ is given by
\be
\frac{S}{N} = -\frac{\partial (F/N)}{\partial T} \biggr|_{N,V}
= - \frac{\partial (f/\rho)}{\partial T} \biggr|_{N,V} \,,
\ee
where $f=F/V$ is the free-energy density. The chemical potential 
is given by $\mu = \partial_N F |_{T,V}$ and the pressure $P$ follows
from
\be
P = \mu \, \rho - f = \frac{N^2}{V} \, \frac{\partial (F/N)}{\partial N}
\biggr|_{T} = \rho^2 \, \frac{\partial (f/\rho)}{\partial \rho}
\biggr|_{T} \,.
\ee
Finally, the energy per particle is obtained from $E/N = F/N + T \, (S/N)$.

\subsection{Hartree-Fock NN and 3N diagrams}

The first-order $\vlowk$ contribution, (1,NN) in Fig.~\ref{diags}, 
is given by
\be
\frac{\ohfnn}{V} = \frac{1}{2} \: \tr_{\sigma_1, \sigma_2} 
\int \frac{d{\bf k}_1}{(2\pi)^3}
\int \frac{d{\bf k}_2}{(2\pi)^3} \: n_{k_1} \, n_{k_2} \,
\langle 1 2 \, | \, \vlowk \, (1-P_{12}) \, | \, 1 2 \rangle \,,
\ee
where the trace is over the spins of the two neutrons and $P_{12}$ 
denotes the exchange operator for spin and momenta of nucleons 
$1$ and $2$. Note that the momentum-conserving delta function
is not included in the NN matrix elements.

In neutron matter, the effect of 3N interactions is expected to be 
smaller than in symmetric matter, since
the Pauli principle prevents three neutrons from interacting in a relative
S-state. In the evaluation of the first-order 3N diagram, (1,3N) in
Fig.~\ref{diags}, we follow
Ref.~\cite{Vlowknucmatt}. At the HF level
only the $c_1$ and $c_3$ terms of the long-range $2 \pi$-exchange 
part contribute:
\begin{align}
\frac{\ohfnnn}{V} &= \frac{g_A^2}{4 f_{\pi}^2} \: 
\int \frac{d{\bf k}_1}{(2\pi)^3} \int \frac{d{\bf k}_2}{(2\pi)^3}
\int \frac{d{\bf k}_3}{(2\pi)^3} \: n_{k_1} \, n_{k_2} \, n_{k_3} \,
f_{\text{R}}^2(p,q) \nonumber \\[2mm]
&\times \biggl[ - \frac{4 c_1 m_\pi^2}{f_\pi^2} \biggl( 2 \,
\frac{{\bf k}_{12} \cdot {\bf k}_{23}}{(k_{12}^2 + m_\pi^2)
(k_{23}^2 + m_\pi^2)} + 2 \, \frac{k_{12}^2}{(k_{12}^2 + m_\pi^2)^2} \biggl)
\nonumber \\[2mm]
&+ \frac{2 c_3}{f_\pi^2} \biggl( 2 \,
\frac{({\bf k}_{12} \cdot {\bf k}_{23})^2}{(k_{12}^2 + m_\pi^2)
(k_{23}^2 + m_\pi^2)} - 2 \, \frac{k_{12}^4}{(k_{12}^2 + m_\pi^2)^2} \biggl)
\biggr] \,,
\end{align}
where $g_A = 1.29$, $f_\pi = 92.4 \mev$, $m_\pi = 138.04 \mev$ and
${\bf k}_{ij} = {\bf k}_i - {\bf k}_j$.
As discussed in Ref.~\cite{Vlowk3N}, we use the $c_i$ constants 
extracted by the Nijmegen group in a partial wave analysis with 
chiral $2 \pi$-exchange~\cite{const}:
$c_1 = -0.76 \gevi$ and $c_3 = -4.78 \gevi$, where the dominant
contribution is due to $c_3$.
The low-energy constants $c_i$ are within errors consistent with the
determination from $\pi$N data~\cite{Ulf}, but at present $c_3$ has a
large theoretical uncertainty $\approx 25 \%$, which is not included
in our error bands (see however Fig.~\ref{energyT0}).
For the 3N contribution,
we have the regulator~\cite{Vlowk3N}
\be
f_{\text{R}}(p,q) = \exp \biggl[ - \biggl( \frac{p^2+3 q^2/4}{\lm^2}
\biggr)^4 \biggr]
\ee
where $p$ and $q$ are 
Jacobi momenta. Based on the nuclear matter results of 
Ref.~\cite{Vlowknucmatt}, we expect that the $c_3$ term is 
repulsive and the dominant part of the 3N contribution, and that
the $c_1$ term is small.

\subsection{Second-order anomalous and normal diagrams}

The second-order anomalous contribution, (2,$a$) in Fig.~\ref{diags},
is given by
\begin{align}
\frac{\oa}{V} &= - \frac{1}{2 T} \, \biggl( \, \prod_{i=1}^{3} \,
\tr_{\sigma_i} \int \frac{d{\bf k}_i}{(2\pi)^3} \, \biggr) \,
n_{k_1} \,  n_{k_2} \, (1-n_{k_2}) \, n_{k_3} \nonumber \\[2mm]
&\times \langle 1 2 \, | \, \vlowk \, (1-P_{12}) \, | \, 1 2 \rangle
\langle 2 3 \, | \, \vlowk \, (1-P_{12}) \, | \, 2 3 \rangle \,.
\end{align}
We note that in the HF approximation, all tadpole self-energy
insertions, including the anomalous diagram (2,$a$), are included 
in the HF mean-field. Hence, in a loop expansion around the HF
solution, the first anomalous diagrams are of fourth order and
involve two second-order self-energy insertions in place of the
tadpoles in diagram (2,$a$).
As discussed above, the present approach is equivalent to a
loop expansion around the HF energy
for a momentum-independent contact interaction. It follows that
the contribution from the square bracket in Eq.~(\ref{eq:F}) is small, 
although the anomalous diagram (2,$a$) is significant. 
The reasons are:
First, at zero temperature the square bracket vanishes,
and therefore the contribution is small at low temperatures,
and second, at finite temperature it is non zero only due to the 
weak momentum dependence of the HF self-energy in
neutron matter.

The second-order normal diagram, (2,$n$) in Fig.~\ref{diags}, reads
\begin{align}
\frac{\on}{V} &= - \frac{1}{8} \, \biggl( \, \prod_{i=1}^{4} \,
\tr_{\sigma_i} \int \frac{d{\bf k}_i}{(2\pi)^3} \, \biggr) \,
(2 \pi)^3 \delta({\bf k}_1 + {\bf k}_2 - {\bf k}_3 - {\bf k}_4) 
\nonumber \\[2mm]
&\times \frac{n_{k_1} n_{k_2} \, (1-n_{k_3}) (1-n_{k_4})
-(1-n_{k_1}) (1-n_{k_2}) \, n_{k_3} n_{k_4}}{\epsilon_{k_3}+\epsilon_{k_4}
-\epsilon_{k_1}-\epsilon_{k_2}} \nonumber \\[2mm]
&\times \bigl| \langle 1 2 \, | \, \vlowk \, (1-P_{12}) \, |
\, 3 4 \rangle \bigr|^2 \,.
\end{align}
Expanding in partial waves and performing the spin traces, we find
\begin{align}
&\sum_{S, M_S, M'_S} \bigl|
\langle {\bf k} \, S M_S \, | \, \vlowk \, (1-P_{12}) \, | \, {\bf k'} \,
S M'_S \rangle \bigr|^2 \nonumber \\[2mm]
&= \sum_L \, P_L(\cos \theta_{{\bf k},{\bf k'}}) \,
\sum_{J,\, l,\, l',\, S} \, \sum_{\tj,\, \tl,\, \tlp} \, (4 \pi)^2 \, 
i^{(l-l'+\tl-\tlp)} \nonumber \\[1mm]
&\times \langle k \, | \,  \vlowk^{J \, l' \, l \, S} \, | \, k' \rangle
\langle k' \, | \,  \vlowk^{\tj \, \tlp \, \tl \, S} \, | \, k \rangle
\, \bigl(1-(-1)^{l+S+1}\bigr) \, \bigl(1-(-1)^{\tl+S+1}\bigr) \nonumber \\[2mm]
&\times \sqrt{(2 l + 1) (2 l' + 1) (2 \tl + 1) (2 \tlp + 1)} \: (2J+1) (2\tj+1)
\: (-1)^{\tl+l'+L} \nonumber \\
&\times \bigl(l \, 0 \, \tlp \, 0  \, | \, L \, 0\bigr)
\bigl(l' \, 0 \, \tl \, 0 \, | \, L \, 0 \bigr)
\biggl\{ \begin{array}{ccc}
l & S & J \\
\tj & L & \tlp \end{array} \biggr\}
\biggl\{ \begin{array}{ccc}
J & S & l' \\
\tl & L & \tj \end{array} \biggr\} \,,
\label{recoupling}
\end{align}
where $\theta_{{\bf k},{\bf k'}}$ is the angle between relative momenta 
${\bf k} = ({\bf k}_1 - {\bf k}_2)/2$ and ${\bf k'} = ({\bf k}_3 - 
{\bf k}_4)/2$. Keeping only $L=0$ in Eq.~(\ref{recoupling}) would result
in an angle average of the Pauli-blocking operator, but all $L \leqslant 6$
are included in our results.

\section{Results}
\label{results}

\begin{figure}
\begin{center}
\includegraphics[scale=0.45,clip=]{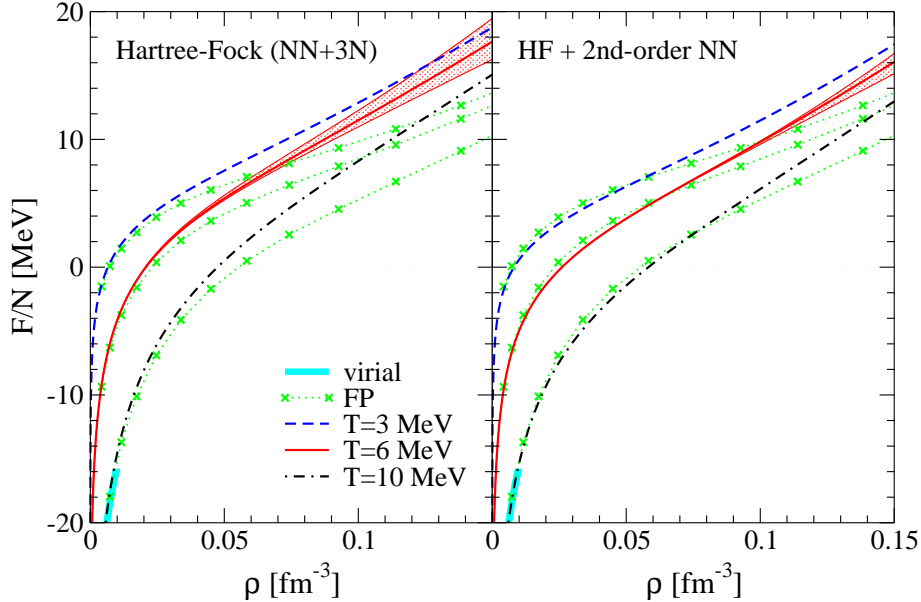}
\end{center}
\caption{The free energy per particle
$F/N$ as a function of density $\rho$. The
left figure gives the first-order NN and 3N contributions with a free 
single-particle spectrum. Second-order anomalous and normal
NN contributions are included 
in the right figure. Our results are compared to the virial equation 
of state (virial)~\cite{vEOSneut} and to the variational calculations 
of Friedman and Pandharipande (FP)~\cite{FP}. The virial curve ends
where the fugacity $z=e^{\mu/T} = 0.5$.}
\label{free_energy}
\end{figure}

We compute the different contributions to the free energy using the 
adaptive Monte Carlo integration routine 
Vegas~\cite{NR}. Our results\footnote{We take the opportunity to 
correct an error
in Ref.~\cite{Proc}, where the 3N contribution had an incorrect factor
in the numerical computation.} 
for the free energy per particle are shown in Fig.~\ref{free_energy}
for temperatures $T=3 \mev, 6 \mev$ and $10 \mev$,
where the low-momentum interaction $\vlowk$ is obtained from the
Argonne $v_{18}$ potential~\cite{AV18} 
for a cutoff $\lm = 2.1 \fmi$. 
The cutoff dependence of the free energy can be used to
provide lower limits for
the theoretical uncertainty in the calculation, since the result 
should be cutoff independent when all relevant contributions are 
included. For the $T=6 \mev$ 
results, we provide error estimates by varying the cutoff over 
the range $\lm=1.9 \fmi$ (lower curve) to $\lm=2.5 \fmi$ (upper curve).
The cutoff dependence of the $T=3 \mev$ and $10 \mev$ results here
and in the following is of similar size.
As expected, the error grows with increasing density. Moreover,
we observe that the equation of state
becomes significantly less cutoff dependent with the inclusion 
of the second-order NN contributions.

In Fig.~\ref{free_energy}, we also compare our results for the 
free energy to the
model-independent virial equation of state~\cite{vEOSneut} and
to the variational calculations of Friedman and 
Pandharipande~\cite{FP} (FP, based on the Argonne $v_{14}$ 
and a 3N potential). 
The virial expansion provides a benchmark for low densities and
high temperatures, where the interparticle separation is large 
compared to the thermal wavelength.
We find a very good agreement with the virial free energy.
Including second-order NN contributions to the HF free energy
brings our results closer to the FP calculations, but this
trend is opposite for other thermodynamic potentials,
see for instance the entropy in Fig.~\ref{entropy}.

The FP results are based on zero-temperature
Fermi-hypernetted-chain correlation functions, with the effective
mass as a finite-temperature variational parameter~\cite{Lagaris}. 
We note that the density of states 
at the Fermi surface is underestimated in variational calculations 
of this type, since the energy dependence of the self energy is properly 
accounted for only when correlation diagrams are 
included~\cite{Mahaux}. In a variational scheme
this can be achieved in correlated basis perturbation 
theory~\cite{FFP,FPS}. This effect is in part included in
the induced interaction, which in neutron matter leads to an
enhancement of the effective mass by $\approx 10 \%$ (see the 
RG results for the Fermi liquid parameter $F_1/3 = m^*/m-1$ 
in Fig.~6 of Ref.~\cite{RGnm}). The enhancement of the effective 
mass near the Fermi surface is reflected in an increase of the 
entropy and the specific heat over the variational result at 
low temperatures~\cite{FPS} (see below).

\begin{figure}
\begin{center}
\includegraphics[scale=0.45,clip=]{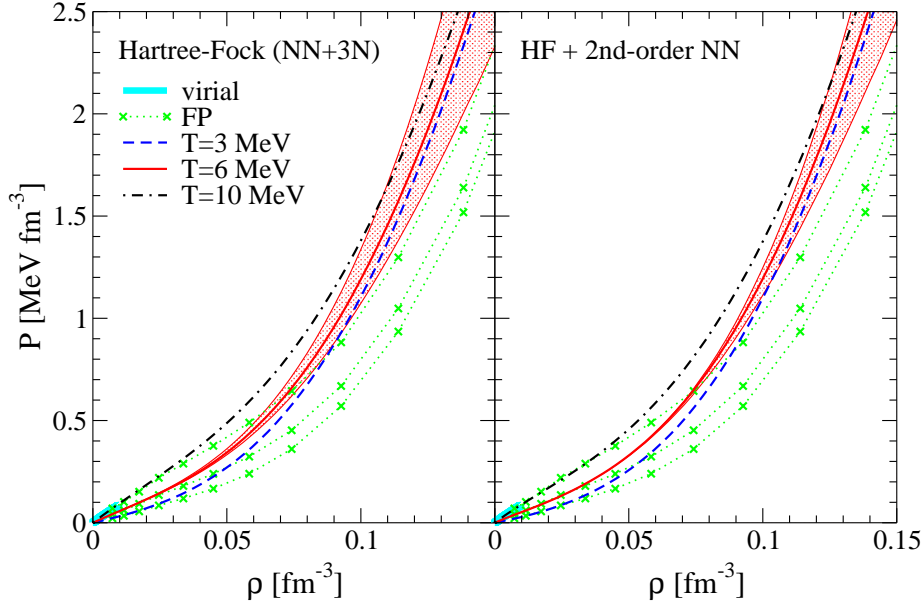}
\end{center}
\caption{The pressure
$P$ as a function of density $\rho$ to first
and second order (for details see 
Fig.~\ref{free_energy}). The different-temperature FP results
are best identified at low density by comparison with our
results.}
\label{pressure}
\end{figure}

\begin{figure}
\begin{center}
\includegraphics[scale=0.45,clip=]{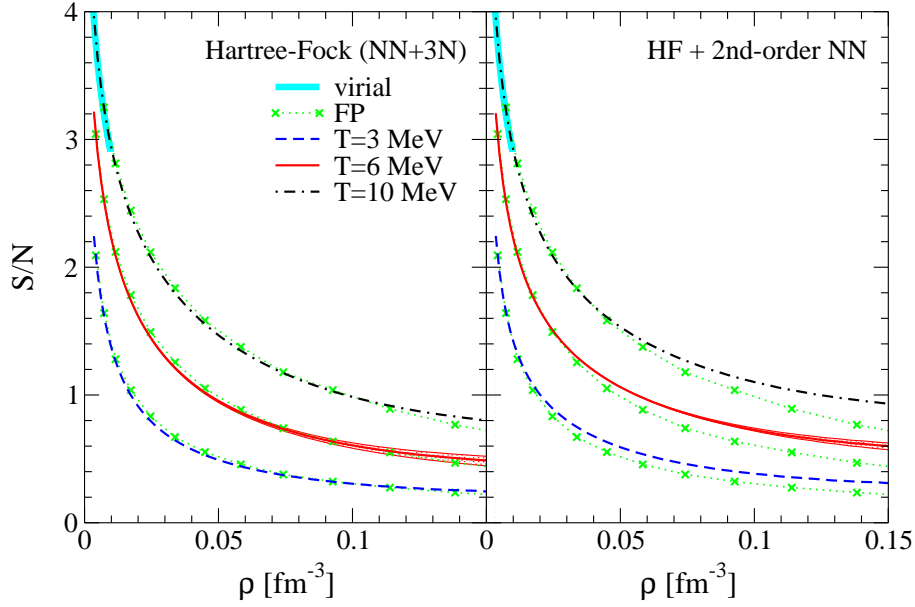}
\end{center}
\caption{The entropy per particle
$S/N$ as a function of density $\rho$ to first
and second order (for details see 
Fig.~\ref{free_energy}).}
\label{entropy}
\end{figure}

\begin{figure}
\begin{center}
\includegraphics[scale=0.45,clip=]{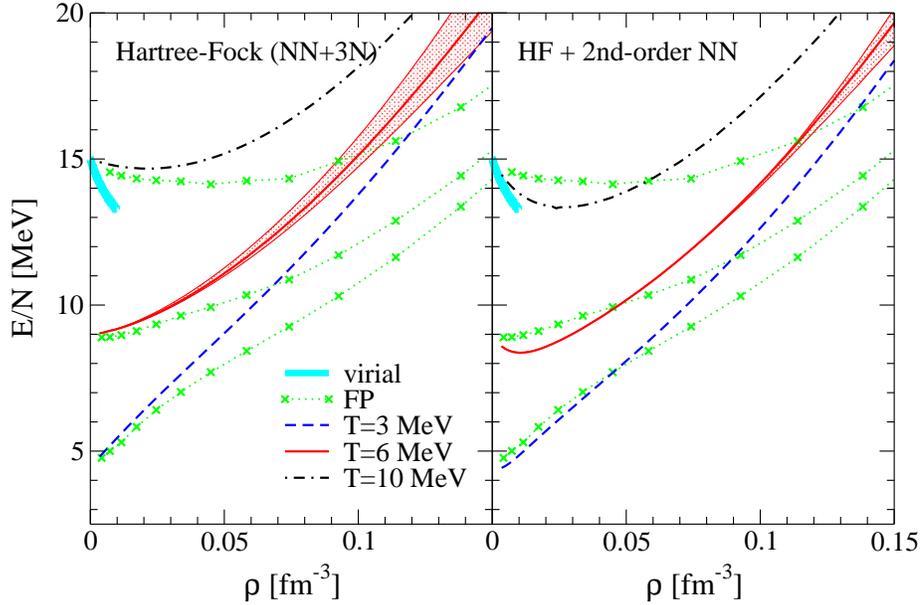}
\end{center}
\caption{The energy per particle
$E/N$ as a function of density $\rho$ to first
and second order (for details see 
Fig.~\ref{free_energy}).}
\label{energy}
\end{figure}

The pressure $P$ and the entropy per particle $S/N$ are shown in
Figs.~\ref{pressure} and~\ref{entropy}. As for the free energy,
we find a very good agreement with the virial equation of state
at low densities,
and the inclusion of second-order contributions significantly
decreases the cutoff dependence. Our results are similar to the
calculations of FP for densities $\rho \lesssim 0.05 \fmiq$.
For higher densities, we find a larger pressure and entropy.
We emphasize that the results are based on different Hamiltonians,
and therefore the comparison has to be taken with care.
However, the dominant source for the difference in the entropy 
is likely due to differences in 
the effective masses, since the entropy density of a low-temperature
Fermi liquid is proportional to the effective mass, $s=m^* k_{\rm F}
T/3$~\cite{BaymPethick}. As discussed above, the variational calculation
underestimates the effective mass at the Fermi surface and
consequently also the entropy at low temperatures.

Our results for the energy per particle are presented
in Fig.~\ref{energy}. As for the free energy, we observe additional 
binding and a significantly reduced cutoff dependence at second 
order. In contrast to the variational calculation of FP~\cite{FP},
the low-density behavior at second order is in good agreement with 
the virial equation of state~\cite{vEOSneut}. This highlights the 
importance of a correct finite-temperature treatment of second
and higher-order contributions. Note that the error
in the virial equation of state (due to the neglected third virial
coefficient) increases with density. This error is not shown in 
Fig.~\ref{energy}, but will be discussed in future work on understanding
the transition from the perturbative to the virial approach.

\begin{figure}
\begin{center}
\includegraphics[scale=0.45,clip=]{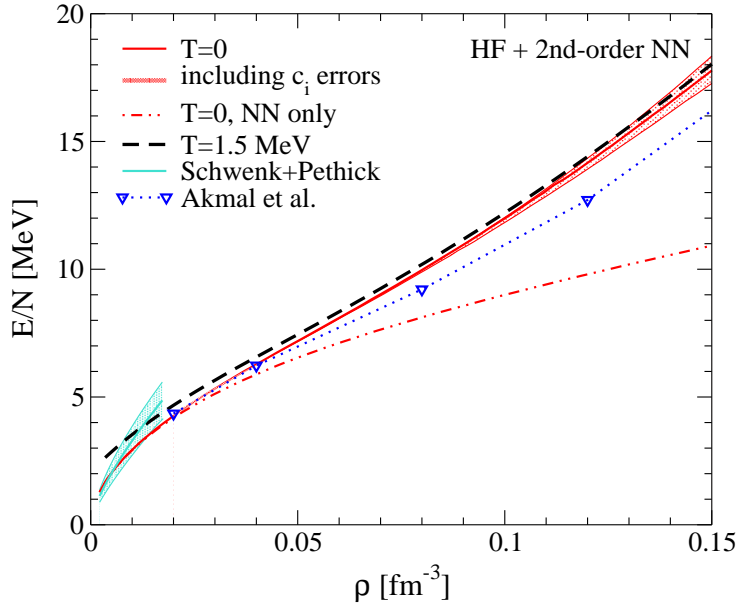}
\end{center}
\caption{The energy per particle $E/N$ as a function of density $\rho$
for $T=0$ and $T=1.5 \mev$. For the $T=0$ results, we provide error
estimates based on the cutoff variation (shaded band) and including the 
uncertainties in the low-energy constants $c_i$ at this level. The upper
and lower limits for the range of $c_i$ values are indicated by the
shaded lines. For comparison we also show the model-independent 
di-fermion EFT results~\cite{dEFT} (Schwenk+Pethick) and the results
of Akmal {\it et al.}~\cite{APR}.}
\label{energyT0}
\end{figure}

A comparison of our low-temperature results to the $T=0$ energy per
particle provides an independent check of our calculations and of the
generalized loop expansion. In Fig.~\ref{energyT0} we show the energy
per particle for $T=1.5 \mev$ for a cutoff $\lm = 2.1 \fmi$ and the
corresponding $T=0$ equation of state. The latter extends the HF
results for neutron matter of Refs.~\cite{RGnm,neutmatt} to include
3N forces and (normal) second-order NN contributions using an
angle-averaged Pauli blocking operator (see Ref.~\cite{Vlowknucmatt}
for details). Except at low densities, where $E/N \to 3/2 T$,
we find that the $T=1.5 \mev$ energy closely follows the zero
temperature results.

In Fig.~\ref{energyT0} we provide an error band for the 
$T=0$ equation of state based on the cutoff variation. The width of 
this band is of the same size as for $T=6 \mev$ in Fig.~\ref{energy}.
At this level of 3N interactions (leading chiral EFT 3N in HF), only
the $2\pi$-exchange part with low-energy constants $c_i$ contribute,
and therefore there are no adjustable 3N parameters. Since the $c_i$
constants are cutoff independent, their uncertainties are not fully captured
by the cutoff variation, and we therefore directly assess how the
presently large uncertainties in $c_i$ propagate to theoretical
uncertainties in the neutron matter equation of state at this level.
The resulting
error estimate in Fig.~\ref{energyT0} is based on $c_1 = -0.9^{+0.2}_{-0.5} 
\gevi$ and $c_3 = -4.7^{+1.2}_{-1.0} \gevi$ from Ref.~\cite{Ulf}.
It is clear that at present the theoretical uncertainties in 3N interactions 
overwhelm the error due to an approximate many-body treatment for
these densities.

We can also compare the $T=0$ energy per particle
at low densities to the the model-independent di-fermion EFT 
results~\cite{dEFT} based directly on the large
neutron-neutron scattering length and effective range. Our results
are consistent with the di-fermion EFT energy per particle within 
errors. Finally, the results of Akmal {\it et al.}~\cite{APR} (based
on the Argonne $v_{18}$ and Urbana IX potential) lie within our
error band as well (including the $c_i$ uncertainties).
The Urbana IX 3N interaction corresponds to the $\Delta$ contribution, 
$c_3^\Delta = -3.83 \gevi$~\cite{BKM}, and therefore results in
less repulsion (with weaker $c_3$). These results show that, 
at present, understanding 3N interactions is a frontier for
nuclear matter at the extremes.

\section{Conclusions}
\label{conclusions}

This work is part of a program to improve the nuclear equation of 
state for astrophysics. One of the central objectives is to
quantify the theoretical uncertainties in the microscopic nuclear 
physics input, and to explore the impact on supernovae 
and neutron stars, for example, through predictions of neutron star
masses and radii.

In this first study of neutron matter, we have computed the 
equation of state at subsaturation densities and temperatures 
$T \leqslant 10 \mev$ based on low-momentum NN and 3N interactions.
We have generalized the perturbative approach~\cite{Vlowknucmatt}
to finite temperature, where the free energy is obtained from a 
loop expansion around the HF energy and the momentum dependence of 
the self-energy is treated perturbatively. Our results include
first-order NN and 3N contributions, as well as anomalous and 
normal second-order diagrams with NN interactions. The pressure, 
entropy and energy were then calculated using standard thermodynamic
relations. While the HF energy is sizable (and non-perturbative
for finite nuclei),
the finite-temperature loop expansion around 
the HF energy seems to be tractable. This is due to a combination
of Pauli blocking~\cite{Vlowknucmatt} and an appreciable effective
range~\cite{dEFT}.

The virial expansion provides a model-independent equation of 
state for nuclear matter at low density and high 
temperature~\cite{vEOSnuc,vEOSneut}, and our perturbative results 
meet this benchmark. This is very promising, since it will enable
us to match the virial equation of state to microscopic calculations
based on NN and 3N interactions at higher densities. The comparison
of our results to the virial energy per particle highlights the 
importance of a correct finite-temperature treatment of second 
and higher-order correlations, which are included only in an average
sense in the variational calculations of Ref.~\cite{FP}. The correct
treatment of thermally-excited low-lying states leads to an
enhancement of the effective mass at the Fermi surface and
consequently to an increase in the entropy, as shown in Fig.~\ref{entropy}.

We have provided simple estimates for the theoretical error by
varying the cutoff in low-momentum interactions. This is a powerful 
tool to assess theoretical errors due to neglected higher-order 
many-body forces and due to an approximate many-body treatment.
We found that the equation of state becomes significantly less 
cutoff dependent with the inclusion of second-order
contributions, and that the cutoff dependence is small for
$\rho \lesssim 0.1 \fmiq$. We note that the errors of the free 
energy are correlated between different temperatures and grow
with increasing density. The first observation implies a relatively
small error in the entropy (obtained by a temperature derivative),
and consequently similar errors for the energy per particle.
The second observation explains the relatively large
error band for the pressure (obtained from a density derivative).
Finally, we have shown that the uncertainties due to the long-range
parts of 3N interactions, the $c_i$ constants, are substantial
and overwhelm the error bands from the cutoff variation at this level. We
conclude that understanding 3N forces is
a frontier in microscopic calculations
of the nuclear equation of state, and furthermore that the possibility of 
estimating theoretical uncertainties is an important step towards
reliable extrapolations to astrophysical conditions.

Future work will include systematic studies of the range of validity,
quantifying an expansion parameter for the loop expansion,
calculations for asymmetric matter, improving the uncertainties
at higher densities, how they propagate to astrophysical observables,
and understanding the transition to the virial expansion.

\begin{ack}
We thank Scott Bogner, Dick Furnstahl and Chuck Horowitz for useful
discussions. AS thanks the GSI Theory Group for the warm hospitality.
This work was supported in part by the Virtual Institute VH-VI-041 of
the Helmholtz Association, by the BMBF projects ANBest-P and BNBest-BMBF
98/NKBF98, and by the Natural Sciences and Engineering 
Research Council of Canada (NSERC). TRIUMF receives federal funding
via a contribution agreement through the National Research Council of Canada.
\end{ack}

\end{document}